\documentstyle[epsf,12pt,psfig,a4wide, amssymb,latexsym,newlfont]{article}
\newcommand{\be}{\begin{equation}}
\newcommand{\ee}{\end{equation}}
\newcommand{\bea}{\begin{eqnarray}}
\newcommand{\eea}{\end{eqnarray}}
\renewcommand\bar[1]{\overline{#1}}

\newcommand\R{\mathbb{R}}
\newcommand\Z{\mathbb{Z}}

\newcommand{\bi}{\bibitem}
\def\sqr#1#2{{\vcenter{\vbox{\hrule height.#2pt
            \hbox{\vrule width.#2pt height#1pt \kern#1pt
                  \vrule width.#2pt}\hrule height.#2pt}}}}
\def\square{
\mathop{\mathchoice{\sqr{20}{25}}{\sqr{12}{15}}{\sqr{8}{10}}{\sqr{4}{5}}}}
\title{The Orbifolds  of $N = 2$ Superconformal Theories with $c = 3$}
\author{\footnotesize Sayipjamal Dulat\\
\footnotesize University of Bonn Physics Institute, Nussallee 12, 53115 Bonn, 
                  Germany }
\date{}
\begin{document}
\maketitle

\begin{abstract}
We construct  
$\Z_M$, $M= 2, 3, 4, 6$ orbifold models of the $N=2$ superconformal
field theories with  central charge $c=3$. Then we check the description
of the  $\Z_3$, $\Z_4$ and $\Z_6$ orbifolds  by  
the $N=2$ superconformal Landau-Ginzburg models with $c=3$, by comparing  
the spectrum of chiral fields, in particular the Witten index $Tr(-1)^F$.
\end{abstract}
\section{Introduction}
The complete understanding of the moduli space of $N=2$ superconformal field 
theories with central charge $c=3$ needs a description of all its orbifold 
theories.
In a non-linear $\sigma$-model description, this concerns two dimensional
tori and their orbifolds. For $\Z_3$, $\Z_4$  and $\Z_6$ orbifolds, 
C.~Vafa and N.~Warner $\cite{vaw}$ made predictions for (chiral, chiral)
and (antichiral, antichiral) fields based
on Landau-Ginzburg descriptions. Apparently, they never had been checked 
explicitely. The moduli spaces of those orbifold theories were obtained in 
$\cite{ler}$.
Landau-Ginzburg descriptions for the three orbifolds, we
 use the superpotentials
$\Phi_1^3 + \Phi_2^3 + \Phi_3^3 
             + 6a\Phi_1\Phi_2\Phi_3$,
$\Phi_1^4 + \Phi_2^4 + a\Phi_1^2\Phi_2^2$,  and  
 $\Phi_1^3 + \Phi_2^6 + a\Phi_1^2\Phi_2^2$, respectively. Note that we are 
interested in one dimensional moduli spaces, such that one needs 
superpotentials with one free parameter a or, in other words, singularities 
of modality one.
Correlation functions for these potentials have been studied in 
\cite{fgls}\cite{ziad}.
Here we calculate the $\Z_M$ orbifold partition functions and check the 
predictions of C.~Vafa and N.~Warner.
For $c=6$ similar calculations have been formulated by T.~Eguchi et al 
$\cite{egu}$. There, charges behave in a simpler way than for $c=3$. When 
fermions are omitted from the $c=3$ superconformal theories, 
one obtains $c=2$ bosonic theories. In this case 
the partition function for the $\Z_2$ orbifold was given in $\cite{ken}$.

The $N=2$ superconformal field theories with  $c=3$ \cite{kent}
are described  by a free chiral scalar superfield containing two real 
bosons or a single complex left (right) boson  
$\varphi^\pm(z) = \varphi^1(z)\pm i\varphi^2(z)$ 
$(\bar{\varphi}^\pm(\bar z) =
        \bar{\varphi}^1(\bar z)\pm i\bar{\varphi}^2(\bar z) )$
(each of $c=1$) and two Majorana-Weyl (MW) fermions or a free complex 
left(right) fermion 
$\psi^\pm(z) =\psi^1(z) \pm i\psi^2(z)$
$(\bar{\psi}^\pm(\bar z) =\bar{\psi}^1(\bar z) \pm i\bar {\psi}^2(\bar z))$
(each of $c=\frac{1}{2}$).
The action for this system may be written as
\be\label{ac}
S=\frac{1}{2\pi}\int\, d^2z (G_{ij}\partial\varphi^i\bar{\partial}\varphi^j +
 B_{ij} \partial\varphi^i\bar{\partial}\varphi^j+
 \psi^-\bar{\partial}\psi^+ +\psi^+ \bar{\partial}\psi^-)  .
\ee
In string theory language, 
 this action corresponds to the superstring compactification 
on a two dimensional torus $T^2 = \R^2/{\Lambda}$. For  
the two dimensional lattice $\Lambda$, we use 
a basis $\{e_i\}\in \R$ $(i = 1,2)$. 
The action (\ref{ac}) depends on four real parameters or moduli,
the constant symmetric metric   
 $G_{ij} = \frac{1}{2}e_ie_j$ on 
$T^2$, and  the antisymmetric tensor field
 $B_{ij}= -B_{ji}$. It has $N=2$ 
superconformal symmetry. Directly from the action, we can determine the 
generators of the $N=2$ superconformal algebra,
the stress-energy tensor $T(z)$, 
its super partners $Q^i(z) = Q^1(z) \pm i Q^2(z)$ ($i=1,2$), 
and the $U(1)$ current $J(z)$ with conformal 
dimensions $h$ equal to $2$, $3/2$, and $1$, respectively

\bea\label{schar}
T(z) & = & -\frac{1}{2}\partial\varphi^-(z)\partial\varphi^+(z) -
           \frac{1}{4}\psi^-\partial \psi^+(z) - 
           \frac{1}{4}\psi^+(z)\partial\psi^-(z)\nonumber\\
Q^\pm(z)&=&\psi^\mp(z)\partial\varphi^\pm(z), \hspace{6mm} 
J(z)=\frac{1}{2}\psi^-(z)\psi^+(z)=\frac{i}{2}\varepsilon^{ij}
            \psi^i(z)\psi^j(z).
\eea
Similar relations hold for the antiholomorphic (right moving)
generators of the $N=2$
superconformal algebra.
They have the Laurent expansions 
$$
T(z) = \sum_{n=-\infty}^{+\infty}L_nz^{-n-2},\hspace{3mm}
Q^i(z) = \sum_{n=-\infty}^{+\infty}Q_r z^{-r-3/2},\hspace{3mm}
J(z) = \sum_{n=-\infty}^{+\infty}J_nz^{-n-1}, 
$$
and satisfy $N=2$ superconformal algebra that can be found in  
\cite{kent}\cite{ss}.
There are three different $N=2$ superconformal algebras, namely 
Ramond (R) (or periodic (P)), Neveu-Schwarz (NS) (or antiperiodic (A)) and 
twisted (T) algebras which correspond 
to different ways of choosing boundary conditions on the cylinder. 
Whatever boundary condition we choose the Virasoro generator $L_n$
is always integrally moded, because the bosonic stress-energy 
tensor is always periodic on the cylinder. For the Ramond (R) 
algebra, $J_n$ and 
$Q_r^i$ are integrally moded, i.e. n and r run over integral values.
For the Neveu-Schwarz (NS) algebra, $J_n$ are integrally moded, 
$Q_r^i$ are 
half integrally moded, i.e. r run over half integral values. 
The twisted (T) algebra has integer modes for $Q_r^1$, half integer modes 
for $J_n$ and $Q_r^2$.

A field satisfying 
$h=\pm\frac{q}{2}$  
is a left chiral or left antichiral primary field.
(Similarly, a field satisfying $\bar h=\pm\frac{\bar q}{2}$
is a right chiral or right antichiral primary field).
Note that the 
fermionic fields $\{\psi^{\pm}(z), \bar{\psi}^{\pm}(\bar z)\}$ 
all satisfy the above condition since they have
charge $\pm 1$ and 
conformal dimension $\frac{1}{2}$ for both the left movers and right 
movers.
The left primary chiral 
fields  are $\{1, \psi^+(z)\}$; the right chiral primary fields are  
$\{1,\bar{\psi}^+(\bar z)\}$. The left and right antichiral primary fields are 
obtained from these by complex conjugation.
Note that the conformal dimensions and $U(1)$ charges 
of an unique highest left-right  
chiral or antichiral primary field are  
$(h, \bar h) = (c/6, c/6)= (\frac{1}{2}, \frac{1}{2})$ and 
$(q, \bar q) = (\pm c/3,\pm c/3)= (\pm 1, \pm 1)$, respectively (here $c=3$).

In general for $N=2$ superconformal theories, 
there are four types of rings \cite{llw} arising from the 
various combinations of left-right chiral and left-right antichiral fields.
We denote these rings by $(c,c)$, $(a,a)$, $(a,c)$, $(c,a)$. They are pairwise
conjugate. For the $\Z_M$, $M\in \{3,4,6\}$, orbifolds of 
$N=2$ superconformal theories with $c=3$, and for 
$N=2$ superconformal Landau-Ginzburg models,
one obtains only $(c, c)$ and its conjugate $(a, a)$ rings. For such models,  
the $(a,c)$ and $(c, a)$ rings are trivial and consist only of the identity 
operator. We shall see this point explicitely in the discussion of $\Z_M$ 
orbifolds and Landau-Ginzburg models.

The basic linearly independent elements of the 
$(c,c)$ ring of the $N=2$ superconformal field theory with 
$c=3$ is given by 
\be\label{ccr}
{\mathcal{R}}_{(c,c)} = \{1,\; \psi^+(z),\;\bar{\psi}^+(\bar z),\; 
                            \psi^+(z)\bar{\psi}^+(\bar z)\}  .
\ee
Similarly, for the (a,c) ring  one has 
\be\label{acr}
{\mathcal{R}}_{(a,c)}= \{1,\; \psi^-(z),\;\bar{\psi}^+(\bar z),\; 
                            \psi^-(z)\bar{\psi}^+(\bar z)\}  .
\ee
The elements of the two other rings ${\mathcal{R}}_{(a,a)}$ and 
${\mathcal{R}}_{(c,a)}$ are obtained from ${\mathcal{R}}_{(c,c)}$ and 
${\mathcal{R}}_{(a,c)}$ by complex conjugation. 

The conformal dimensions and $U(1)$ charges of the  ground states of 
Ramond sector are $(h,\bar h)= (c/24,c/24) = (1/8,1/8)$ and 
$(q, \bar q) = (\pm 1/2, \pm 1/2)$, which also contribute to
the Witten index $Tr(-1)^F$ 
\cite{wit}.
The operator $(-1)^F$, where $F = F_L + F_R$, and $F_L$, $F_R$  are left-right
moving fermion numbers, defined to anticommute with all the fermionic 
operators  $(-1)^F\psi(z) = -\psi(z)(-1)^F$,
and to commute with all the bosonic operators
$(-1)^F\varphi(z) = \varphi(z)(-1)^F$, as well as to satisfy
$((-1)^F)^2$. It can be defined in terms of zero mode $U(1)$ current as 
$$
(-1)^F = e^{\pi i(J_0 - \bar{J}_o)}.
$$

It is well known that one can connect the Neveu-Schwarz sector
 to the Ramond 
sector by spectral flow \cite{ss} operation. 
It is the continous transformation and has the following form
\begin{eqnarray*}
L_n^\eta & =&  L_n + \eta J_n + \frac{c}{6}\eta^2\delta_{n,0} \nonumber\\
J_n^\eta & =& J_n + \frac{c}{3}\eta\delta_{n,0} \nonumber\\
Q_r^{\pm \eta} & =& Q_{r \pm \eta}^\pm.
\end{eqnarray*}
The $\eta$ twisted operators $L_n^\eta$, $Q_r^{\pm \eta}$ and 
$J_n^\eta$ still satisfy the $N=2$ superconformal algebra for an arbitrary 
value of the parameter $\eta$. In particular,
the zero mode eigenvalues $h$ of $L_0$ and  $q$ of $J_0$ are changed by 
spectral flow as 
\be\label{spectralflow}
h_\eta = h + \eta q + \eta^2 \frac{c}{6}, \hspace{10mm}
q_\eta = q +  \eta \frac{c}{3} .
\ee 
By (\ref{spectralflow}) 
with flow parameter $\eta = \frac{1}{2}$, the ground states of 
Ramond sector with conformal dimension $(h,\bar h)= (1/8,1/8)$ and charge  
$(q, \bar q) = (\pm 1/2, \pm 1/2)$
flow to 
the Neveu-Schwarz chiral primary fields 
with conformal dimension  $(h,\bar h) = (1/2,1/2)$
and charge $(q, \bar q) = (+1, +1)$, or $(h,\bar h) = (q, \bar q) = (0,0)$.
The flow between the NS and NS as well as R and R can be obtained by the flow 
parameter $\eta = 1$.
Besides, under the left-right symmetric spectral flow, $q - \bar q\in \Z$ 
does not change. Thus the Witten's index \cite{ler} is
\bea\label{R}
Tr(-1)^F &=& Tr_R\left[(-1)^{J_0 - \bar J_0} q^{L_0 - \frac{c}{24}}
       \bar {q}^{\bar {L}_0 - \frac{c}{24}}\right]\nonumber\\
&=& Tr_{\mathcal{H}_\eta} \left[(-1)^{J_0^\eta - \bar J_0^\eta} 
               q^{L_0^\eta - \frac{c}{24}}
         \bar {q}^{\bar {L}_0^\eta - \frac{c}{24}}\right]\nonumber\\
&=&Tr_{NS}\left[(-1)^{J_0 - \bar J_0} q^{L_0 - \frac{1}{2}J_0}
                      \bar {q}^{\bar {L}_0 - \frac{1}{2} \bar{J}_0}\right] 
= \sum_{\mathcal{R}} e^{i\pi(q - \bar q)} ,
\eea
where the ${\mathcal{H}_\eta}$ in the second line 
is the Hilbert space of states 
which is twisted by the parameter $\eta$. 
The $\mathcal {R}$  in the last line denotes the chiral ring. 
First line implies that the ground state of the Ramond sector gives 
nonvanishing contribution to the Witten index. The second line is obtained by
applying the spectral flow to the first line. By setting 
$\eta = \frac{1}{2}$ one can flow from Ramond sector to 
the Neveu-Schwarz sector. 
(Note that $J_0^\eta - \bar J_0^\eta =J_0 -\bar J_0$).
Thus  the Witten index receives 
contributions from either the ground states of Ramond sector or the 
chiral 
primary states of  Neveu-Schwarz sector. 
The only difference between the charges of 
the NS chiral primary states and that of the Ramond 
ground states is $\frac{c}{6}$.

The Poincar\'{e} polynomial \cite{llw} is
\be\label{care}
P(t, \bar t ) = Tr_{\mathcal R}t^{J_0\bar J_0}, 
\ee
which  satisfies a duality  relation  
$P(t, \bar t) = (t\bar t)^{1/3}P(1/t, 1/\bar t)$.
Here $t$ and $\bar t$ can be regarded  as an independent variables.
By (\ref{R}), (\ref{care}) and (\ref{ccr}), the Witten index and 
the Poincar\'{e} polynomial are 
\be\label{ppn2} 
Tr(-1)^F = 0, \hspace{7mm}  P(t, \bar t )_{(c,c)} = 1 + t  + \bar t + t\bar t.
\ee
One notes that the 
Poincar\'{e} polynomial (\ref{ppn2}) and ring structure for 
 $(c,c)$ and $(a,c)$ primary fields 
are isomorphic. However, this is not true in general.

The partition function for the $N=2$ superconformal theories 
with $c=3$ is constructed 
by tensoring the theory of a complex free boson 
defined on a 2-dimentional torus $T^2$ 
in the presence of constant background fields, 
with the theory of a single complex Dirac fermion, namely
$$\label{zc3}
Z(\tau,\rho,z) := Z(\tau,\rho,\sigma)Z_{Dirac}(\sigma, z),
$$
In the following we briefly discuss how the explicit expression of 
$Z(\tau,\rho,z)$
can be formulated.
The $Z(\tau,\rho,\sigma)$ 
is the modular invariant partition function for two real 
boson compactified on the two dimensional torus \cite{fms}
\be\label{Zto}
Z(\tau,\rho):=Z(\tau,\rho,\sigma)
             = tr{q^{L^b_0 -\frac{1}{12}} \bar{q}^{\bar{L}^b_0 -\frac{1}{12}}}
                      =
\frac{1}{\left|\eta^2(\sigma)\right|^2}\sum_{ {n_1, m_1 \atop n_2, m_2}
						} 
	q^{\frac{p^2}{2}}\bar q^{\frac{\bar {p}^2}{2}}, 
\ee
where  
$q = e^{2\pi i\sigma}$ , $\sigma = \sigma_1 + i\sigma_2$ parametrizes  
the world sheet torus, 
and $\eta(\sigma)$ is the Dedekind eta function
defined as
$$
\eta(\sigma) = q^\frac{1}{24}\prod_{n=1}^{\infty} (1- q^n). 
$$
The Virasoro zero mode operators for the bosons in (\ref{Zto}) are given by  
\be\label{vir}
L^b_0 = \sum_{n>0}\alpha_{-n}^i \alpha_n^i  + \frac{1}{2}p^2, \hspace{6mm}
\bar{L}^b_0 = \sum_{n>0} \bar{\alpha}_{-n}^i \bar{\alpha}_{n}^i + 
   \frac{1}{2}\bar{p}^2.
\ee
The left-right moving zero mode 
momentum  $p$ and $\bar{p}$ in (\ref{Zto}) 
are defined as  
\be\label{mo}
(p, \bar p) := \left(n_i e^{* i} + e^{*i} B_{ji} m^j + \frac{1}{2} e_j m^j\;,
              \; n_i e^{* i} + e^{*i} B_{ji} m^j - \frac{1}{2} e_j m^j\right),
\ee
where
$\{e^*_i\}$ are basis vectors for the dual lattice $\Lambda^*$ of $\Lambda$,
which satisfies $e_ie^*_j = \delta_{ij}$ such that 
$e^{*i}e^{*j} = \frac{1}{2}G^{ij}$; 
the integers $n_i$ and $m_i$ are the momentum 
and winding numbers. 
The action of $L^b_0$ and $\bar{L}^b_0$ in (\ref{vir}) on the ground state
 $|m_1, m_2, n_1, n_2\rangle$, which is labeled by 
the momentum and winding numbers, is given by
$$
L^b_0|m_1, m_2, n_1, n_2\rangle = \frac{1}{2}p^2|m_1, m_2, n_1, n_2\rangle,
\hspace{6mm}
\bar{L}^b_0|m_1, m_2, n_1, n_2\rangle =
          \frac{1}{2}\bar p^2|m_1, m_2, n_1, n_2\rangle .
$$
%
where we have used 
          $\alpha_n^i |m_1,m_2, n_1, n_2\rangle=0$ 
    and 
                       $\bar{\alpha}_m^j|m_1,m_2, n_1, n_2\rangle = 0$  
    for
$n>0$, $m>0$.
It is well known \cite{narain} that 
the momenta in 
(\ref{mo}) form four dimensional Lorentzian lattice with scalar product  
$(p, \bar p)\cdot (p^\prime, \bar{p}^\prime) = 
      (p\cdot p^\prime-\bar p\cdot\bar{p}^\prime)$, which is
even (because $p^2 - \bar{p}^2 =
2m^i n_i \in {2\Z}$ ) and self-dual ( because $\Lambda =\Lambda^*$).
 From (\ref{mo}), we easily write 
\be\label{mom}
p^2(\bar{p}^2) = \frac{1}{2} n_i n_jG^{ij} + n_i m_j B_{jl} G^{il} 
                  \pm n_i m_i  + \frac{1}{2} m_i m_j(
                        G_{ij} +B_{jk}B_{il} G^{kl} ).
\ee
In the two dimensional case, 
it is convenient to group the four real parameteres
($G_{11}$, $G_{12}$, $G_{22}$, and $B_{12}$)
in terms of two parameters $\tau$ and $\rho$ in the upper complex half plane 
as follows
$$
\tau = \tau_1 + i\tau_2 = \frac{G_{12}}{G_{22}} + 
                               i\frac{\sqrt G}{G_{22}},\hspace{10mm}
\rho = \rho_1 + i\rho_2 = B_{12} + i\sqrt G.
$$
Here 
$\tau$ represents the complex structure of the target space 
torus $T^2$, 
and $\rho$ is its complexified K\"ahler structure; both take values on 
the complex upper half plane; $G = det(G_{ij})$.
Now we write $(\ref{mom})$ in terms of $\tau$ and $\rho$ in the following form
\begin{eqnarray*}
 p^2 & = &\frac{1}{2\tau_2\rho_2}
                  {|n_1 - \tau n_2 - \rho (m_2 + \tau m_1)|}^2\\
\bar{p}^2 & = & \frac{1}{2\tau_2\rho_2}
                {|n_1 - \tau n_2 - \bar{\rho} (m_2 + \tau m_1)|}^2.
\end{eqnarray*}
Finally, torus partition function (\ref{Zto}) takes the  
form
\be\label{tor}
Z(\tau,\rho) =
\frac{1}{\left|\eta^2(\sigma)\right|^2}\sum_{ {n_1, m_1 \atop n_2, m_2}
						} 
           q^{\frac{1}{4\tau_2\rho_2}
                  {|n_1 - \tau n_2 - \rho (m_2 + \tau m_1)|}^2
       }
\bar{q}^{\frac{1}{4\tau_2\rho_2}
                {|n_1 - \tau n_2 - \bar{\rho} (m_2 + \tau m_1)|}^2}.
\ee  
If $\tau_1 =\rho_1 = 0$ (or $G_{12} = B_{12} = 0$), then the torus 
partition function (\ref{tor}) is the product of two  circle
 partition functions \cite{gin} at $c=1$
with radius $r_1=\sqrt{G_{22}}=\sqrt{\rho_2/\tau_2}$ 
and $r_2 = \sqrt{G_{11}}=\sqrt{\tau_2\rho_2}$
$$
Z(\tau_2, \rho_2) = Z^{c=1}(\sqrt{\rho_2/\tau_2})Z^{c=1}(\sqrt{\tau_2\rho_2}).
$$
The partition function for the Dirac fermion 
can be constructed  
by taking equal spin structures for the left and right fermions \cite{gin}
\bea\label{Zdi}
Z_{Dirac}(\sigma, z)   &=& 
 tr{q^{L^f_0 -\frac{1}{24}} \bar{q}^{\bar{L}^f_0 -\frac{1}{24}}}
    y^{J_0}\bar{y}^{\bar{J}_0}\nonumber\\
             &=&
\frac{1}{2}\left(\left|\frac{\vartheta_1(z, \sigma)}{\eta(\sigma)}\right|^2
          +\left |\frac{\vartheta_2(z, \sigma)}{\eta(\sigma)}\right |^2  
          +\left |\frac{\vartheta_3(z, \sigma)}{\eta(\sigma)}\right |^2
       +\left |\frac{\vartheta_4(z, \sigma)}{\eta(\sigma)}\right |^2 \right),
\eea
where $y = e^{2\pi i z}$.
Since the fermionic theory split into Neveu-Schwarz and Ramond sector 
the Virasoro zero mode generator for the Dirac fermions in (\ref{Zdi})
is given by 
$$
L^f_0 =  \sum_{n>0} n d_{-n}^i d_n^i
\hspace{6mm} n{\in\Z +\frac{1}{2}}\hspace{6mm} (NS),\hspace{6mm}
L^f_0 =  \sum_{n>0} n d_{-n}^i d_n^i + \frac{1}{8}
\hspace{6mm} n\in\Z\hspace{7mm} (R).
$$
Similar relation is true for the right moving component.
The classical Jacobi 
theta functions $\vartheta_i(z,\sigma)$, $i\in\{1,2,3,4\}$ in (\ref{Zdi})
are defined  
in terms of sums and products as
\begin{eqnarray*}\label{gejac}
\theta_1(z, \sigma) & = & -i\sum_{n=-\infty}^{\infty} 
                    (-1)^n q^{\frac{1}{2}(n-\frac{1}{2})^2} y^{n-\frac{1}{2}}
                          =\;\;
 -iy^{\frac{1}{2}} q^{\frac{1}{8}}\prod_{n=1}^{\infty}
                       (1- q^n) (1-y q^n)(1- y^{-1} q^{n -1}) 
                                    \\
\theta_2(z,\sigma) & = & \sum_{n=-\infty}^{\infty} 
                    q^{\frac{1}{2}(n-\frac{1}{2})^2} y^{n-\frac{1}{2}}
                  \hspace{15mm} =\;\;
 y^{\frac{1}{2} } q^{\frac{1}{8}} \prod_{n=1}^{\infty} 
              (1- q^n) (1+ y q^n)(1+ y^{-1} q^{n -1}) 
                            \\
\theta_3(z,\sigma) & = & \sum_{n=-\infty}^{\infty} 
                    q^{\frac{n^2}{2}} y^n
                \hspace{27mm} =\;\;
                \prod_{n=1}^{\infty}(1- q^n)
           (1+ y q^{n - \frac{1}{2}})(1+ y^{-1} q^{n -\frac{1}{2}}) 
                            \\
\theta_4(z,\sigma) & = & \sum_{n=-\infty}^{\infty} 
                    (-1)^n q^{\frac{n^2}{2}} y^n
                      \hspace{17mm} =\;\;
                \prod_{n=1}^{\infty}
            (1- q^n)(1- y q^{n - \frac{1}{2}})(1- y^{-1} q^{n -\frac{1}{2}}).
\end{eqnarray*}
Partition function for the $N=2$ superconformal theories with $c=3$
is thus given as 
\bea\label{zc3}
Z(\tau,\rho,z) :&=& Z(\tau,\rho) Z_{Dirac}(\sigma, z)\nonumber\\
                      &=&
               \frac{1}{\left|\eta^2(\sigma)\right|^2}
         q^{\frac{1}{4\tau_2\rho_2}
                  {|n_1 - \tau n_2 - \rho (m_2 + \tau m_1)|}^2}
   \bar{q}^{\frac{1}{4\tau_2\rho_2}
                {|n_1 - \tau n_2 - \bar{\rho} (m_2 + \tau m_1)|}^2}\times
               \nonumber\\
&&
\frac{1}{2}\left(\left|\frac{\vartheta_1(z, \sigma)}{\eta(\sigma)}\right|^2
          +\left |\frac{\vartheta_2(z, \sigma)}{\eta(\sigma)}\right |^2  
          +\left |\frac{\vartheta_3(z, \sigma)}{\eta(\sigma)}\right |^2
       +\left |\frac{\vartheta_4(z, \sigma)}{\eta(\sigma)}\right |^2 \right).
\eea
\section{General Prescription for $\Z_M$ Orbifold Construction}
In this section  we will give the general procedure for the construction of 
the $\Z_M$ orbifolds. 
In fact there 
are not many two dimensional $\Z_M$ orbifolds, because the order M rotation 
must be an automorphism of some two dimensional lattice; therefore $\Z_M$ must
have order $M=2,3,4$, and $6$. The $M=3$ and $M=6$ require the hexagonal 
lattice $(\tau = e^{2\pi i/3})$; 
$M=4$ requires a square lattice $(\tau = i)$.
Under the $\Z_M$ symmetry bosonic  fields  and its modes $\alpha^{\pm}_n$
 transform as 
\be\label{botrans}
(g^k\varphi)^{\pm}(z) = e^{\pm \frac{2\pi i k}{M}} \varphi^{\pm}(z),
                    \hspace{6mm}
 g^k\alpha^{\pm}_n g^{-k} 
     =  e^{\pm \frac{2\pi i k}{M}}\alpha^{\pm}_n,
           \hspace{4mm} k= 1,2,\cdots, M-1.
\ee
Since we want to discuss superconformal orbifold theories, we should include 
the worldsheet fermion $\psi$'s as well. They transform as
\be\label{fertrans}                                         
(g^k\psi)^{\pm}(z) = e^{\pm \frac{2\pi i k}{M}} \psi^{\pm}(z),\hspace{6mm} 
g^k d^{\pm}_n g^{-k}
                 = e^{\pm \frac{2\pi i k}{M}} d^{\pm}_n,
   \hspace{4mm} k= 1,2,\cdots M-1.
\ee
In fact this is also required by the $N=2$ superconformal invariance.
The $\Z_M$ rotations are the symmetries both the action 
$(\ref{ac})$ and  $N=2$ world sheet supersymmetry generators $(\ref{schar})$.
Thus
the two dimensional $N=2$ superconformal  orbifold models $T^2/{\Z_M}$ may be 
constructed by identifying points of the two-dimensional torus $T^2$ 
under the symmetry group $\Z_M$.

Let $\widetilde{\mathcal H}$ be the Hilbert space of an orbifold theory. 
It has  two sectors, namely untwisted and twisted sector, i.e,  
$\widetilde {\mathcal H} = \widetilde{\mathcal H}_u \oplus 
\widetilde {\mathcal H}_t$.
Let us consider first the untwisted sector of the orbifold theory.
The untwisted Hilbert space will be a subspace of the Hilbert space for the 
$N=2$ theories with  $c=3$. 
In the path integral for the partition function this means that the 
bosonic fields obey periodic boundary 
conditions along the space direction of the torus and twisted periodic 
boundary conditions in time. So 
on an orbifold, the untwisted sector boundary conditions on the bosonic 
field are given as
\bea\label{pboun}
\varphi^+(1) &=&  \varphi^+(0)
                         + 2\pi \Lambda \nonumber\\
\varphi^+(\sigma) &=& g \varphi^+(0)
                        +  2\pi \Lambda,
\eea
where $g\in {\Z_M}$.
 For Ramond or Neveu-Schwarz fermion one has 
\bea\label{fboun}
\psi^+(1) &=&  \pm \psi^+(0) \nonumber\\
\psi^+(\sigma) &=&\pm g \psi^+(0).
\eea
Under the above boundary conditions, the bosonic  field has expansion
\be\label{bmod}
\varphi^+(z) =  q^+ - ip^+\ln z + i\sum_{n \ne 0}
                                   \frac{1}{n}\alpha_{n} ^+ z^{-n},
\ee
for the fermionic field one has  
\bea\label{fmod}
\psi^+(z) & = & \sum_{n} d_n^+ z^{-n}
           \left\{      
                  \begin{array}{ll}
& n\in \Z \hspace{10mm} (R) \\
& n\in {\Z +\frac{1}{2}}\hspace{3,6mm} (NS)
                  \end{array}
          \right..
\eea
The untwisted  Hilbert space $\widetilde{\mathcal H}_u$ decomposes into $\Z_M$ 
invariant and noninvariant space of states.
In oder to construct consistent models, we must project out the group 
noninvariant space of states. 
In the Hamiltonian formalism, group invariant states 
are obtained by insertion of the projection operator 
$P = \frac{1}{|\Z_M|}\sum_{g\in \Z_M}\limits g$  into the trace over
states. Here $|\Z_M|$ is the  number of elements in $\Z_M$ and the sum 
$\sum g$ runs over all elements in $\Z_M$.
Thus the untwisted sector partition function is   
\be\label{unorb}
Z_u = tr_{\widetilde{\mathcal H}_u }
             Pq^{L_0 -\frac{1}{8}}\bar{q}^{\bar L_0 -\frac{1}{8}}
            y^{J_0}\bar{y}^{\bar{J}_0}.
\ee
Here $tr_{\widetilde{\mathcal H}_u }$ denote the trace in the 
untwisted Hilbert space sectors and $L_0 = L^b_0 + L^f_0$.
In the path integral formalism,  projection onto group invariant states 
in the untwisted sector is represented as
$$ 
Z_u = \frac{1}{|\Z_M|}\sum_{g\in {\Z_M}}{\scriptstyle{g}}\square_{1},
$$
where we sum over all possible twistings in the time direction of 
the torus. ${\scriptstyle{g}}\square_{1}\limits$ 
represents boundary conditions on any generic
fields in the theory twisted by  g in the time direction of the torus. 
The partition function of the original model is simply given by 
$Z ={\scriptstyle{1}}\square_{1}\limits$ .

The untwisted sector partition function is not modular invariant; one should 
take into account the contributions of twisted sector Hilbert space of states. 
For each element $h\in \Z_M$ one can construct a twisted Hilbert space 
$\widetilde{\mathcal H}_h$. In the path integral 
description the bosonic field obey 
the twisted boundary conditions
\bea\label{tiwb}
\varphi^+(1) &=&  h\varphi^+(0)
                         + 2\pi \Lambda \nonumber\\
\varphi^+(\sigma) &=& g \varphi^+(0)
                        +  2\pi \Lambda.
\eea
For Ramond or Neveu-Schwarz fermions one has  
\bea\label{tiwf}
\psi^+(1) &=& \pm h\psi^+(0)\nonumber\\
\psi^+(\sigma) &=& \pm g \psi^+(0),
\eea
where h and g are twists on the fields in the space and time direction 
of the torus. The  mode expansion of the bosonic 
field which satisfies the boundary conditions $(\ref{tiwb})$ is
\be\label{tbmod}
\varphi^+(z)  =  q^+_f + i\sum_{n\in {\Z + k/M}} 
            \frac{1}{n} \alpha_n^+ z^{-n}.
\ee
One can not have nonzero momentum or winding number here, since they are not 
consistent with the twisted boundary conditions. 
In this mode expansion $q^+_f$ denote the  fixed points of  
$T^2$ under the  $\Z_M$ symmetry.
The index f labels these fixed points.
The  mode expansion of the fermionic field which satisfies the boundary 
conditions $(\ref{tiwf})$ is
\be\label{tfmod}
\psi^+(z)  = \sum_{n\in {\Z + k/M + 1/2 -s/2}}
               d^+_n z^{-n},\hspace{5mm} k = 1,...M-1,
\ee
where s is equal to zero 
in the Neveu-Schwarz sector, and to one in the Ramond sector.
The twisted Hilbert space 
$\widetilde{\mathcal{H}}_t$  decomposes into $\Z_M$ invariant 
and noninvariant space of states. To construct consistent models, 
we again have to project onto group invariant states. In the Hamiltonian 
formalism, group invariant states are obtained by insertion of the 
projection operator  
$P_h\;:= \frac{1}{|\Z_M|}\sum_{g\in \Z_M:\, [g,h]=0}\limits g$ 
into the trace over states.
In the path integral formalism, projection onto group invariant states 
in the twisted sector is representes as 
$$
Z_t =\frac{1}{|\Z_M|}\sum_{g,h\in \Z_M,\atop h\ne 1, [g,h]=0}
                 {\scriptstyle{g}}\square_{h}\;,
$$
where ${\scriptstyle{g}}\square_{h}\limits$
represents boundary conditions on the fields twisted by g and h in the time 
and space direction of the torus.
Thus the twisted sector partition function  has the form 
\be\label{gtiw}
Z_t =\sum_{h\in \Z_M, h\neq1}\limits tr_{\widetilde{\mathcal H}_h} P_h
                 q^{L_0 -\frac{1}{8}}\bar{q}^{\bar L_0 -\frac{1}{8}}
                 y^{J_0}\bar{y}^{\bar{J}_0}
                    =
\frac{1}{|\Z_M|}\sum_{g,h\in \Z_M,\atop h\ne 1, [g,h]=0}
         {\scriptstyle{g}}\square_{h}.
\ee
In fact, one may obtain 
the twisted sector partition function from (\ref{unorb}) by modular 
transformations $\sigma \rightarrow \sigma + 1$ and 
$\sigma \rightarrow -1/{\sigma}$.
Thus, total modular invariant $\Z_{M}$ 
orbifold partition function is a sum of (\ref{unorb})
and (\ref{gtiw})
\bea\label{gorb}
Z_{\Z_M-orb} & = & 
        \frac{1}{|\Z_M|}\sum_{g\in \Z_M}{\scriptstyle{g}}\square_{1}\;\;
                         + \;\;
\frac{1}{|\Z_M|}\sum_{g,h\in \Z_M, h\ne 1}{\scriptstyle{g}}\square_{h}
\nonumber\\
                  & = & \frac{1}{|\Z_M|}\sum_{g,h\in \Z_M, \atop [g,h]=0}
               {\scriptstyle{g}}\square_{h}
                      \; =\;
\sum_{h\in \Z_M} tr_{\widetilde{\mathcal H}_h} 
                    P_h q^{L_0 -\frac{1}{8}}
                   \bar{q}^{\bar L_0 -\frac{1}{8}}
                   y^{J_0}\bar{y}^{\bar{J}_0}\;,
\eea
where we set $\widetilde{\mathcal H}_1\;:=
       \widetilde {\mathcal H}_u$ and $P_1\;:= P$.
There is no discrete torsion for  the $\Z_M$ orbifolds, since all boxes 
${\scriptstyle{g}}\square_{h}\limits$ are related by modular 
tranformations to a box of type 
${\scriptstyle{g}}\square_{1}\limits$. Mathematically, the discrete torsion 
for a discrete group G 
is obtained from the cohomology $H_2(G)$, which vanishes for $G=\Z_M$
\cite{cum}.

In summary, in order to construct an orbifold model, one first formulates 
the Hilbert space of states on the torus, then one projects onto the group 
invariant 
states, finally one includes twisted sector contributions. For more details 
see ref.$\cite{har}$ $\cite{vafa}$ $\cite{dixon}$. 
\section{The $\Z_2$ Orbifold}
The two dimensional $N=2$ superconformal $\Z_2$ orbifold model $T^2/{\Z_2}$
can be constructed from (\ref{zc3}) 
for arbitrary $\tau$ and $\rho$.
Thus we may now produce another family of theories, i.e.  
$\Z_2$ orbifold superconformal field theories with the same set of moduli 
as the $N=2$ theories with $c=3$ by following the general orbifold 
prescription introduced in section two.
The action of $g\in\Z_2$ on the bosonic Hilbert 
space sectors $|m_1,m_2,n_1,n_2\rangle$ is given by 
\be\label{z2g}
g|m_1,m_2,n_1,n_2\rangle= |-m_1,-m_2, -n_1,-n_2\rangle.
\ee
In the following, we only discuss 
the bosonic part since the sum over the spin structures for the Dirac 
fermion is invariant under $\psi^\pm \rightarrow -\psi^\pm$. 
Under the $\Z_2$ symmetry the untwisted bosonic Hilbert spaces 
$\widetilde{\mathcal H}_u$ decomposes into $g = \pm 1$ 
eigenspaces $\widetilde{\mathcal H}_u =  
         \widetilde{\mathcal H}_u ^+ \oplus \widetilde{\mathcal H}_u^- $ as 
\begin{eqnarray*}
\widetilde{\mathcal H}_u^+ & = & \{ 
   \alpha^+_{-k_1}\cdots\alpha^+_{-k_l}
  \bar{\alpha}^+_{-k_{l+1}}\cdots \bar{\alpha}^+_{-k_{2j}}
         (1 + g)|m_1,m_2,n_1,n_2 \rangle \} \\ 
& & +\;
\{ \alpha^+_{-k_1}\cdots\alpha^+_{-k_l}
  \bar{\alpha}^+_{-k_{l+1}} \cdot\cdot\cdot \bar{\alpha}^+_{-k_{2j +1}} 
(1 - g) |m_1,m_2,n_1,n_2 \rangle \}\\
\widetilde{\mathcal H}_u^- & = & \{ 
   \alpha^+_{-k_1}\cdot\cdot\cdot\alpha^+_{-k_l}
  \bar{\alpha}^+_{-k_{l+1}}\cdots\bar{\alpha}^+_{-k_{2j + 1}}
         (1 + g) |m_1,m_2,n_1,n_2 \rangle 
                       \} \\ 
& & \; + 
\{ \alpha^+_{-k_1}\cdots\alpha^+_{-k_l}
  \bar{\alpha}^+_{-k_{l+1}} \cdot\cdot\cdot \bar{\alpha}^+_{-k_{2j}} 
 (1- g)|m_1,m_2,n_1,n_2 \rangle\},\\
\end{eqnarray*}
where $k_i$ takes positive integer values.
By (\ref{unorb}), untwisted $\Z_2$ orbifold 
 partition function is 
\begin{eqnarray*}
 Z_u =
       (q\bar q)^{-\frac{1}{8}} tr_{\widetilde{\mathcal H}_u}{ 
        \frac {1}{2}(1+g)q^{L_0}\bar{q}^{\bar{L}_0} 
                         y^{J_0} \bar{y}^{\bar{J}_0}}.
\end{eqnarray*}
The first term in the trace 
 is equal to the partition function in $(\ref{zc3})$ since 
there is no twist along the two cycles of the torus. 
The second term in the trace  with $g$ inserted  
receives only contribution from the sector $m_1 = m_2 = n_1 = n_2 = 0$ because 
each state obtained by acting on $(1 + g) |m_1,m_2,n_1,n_2\rangle$ 
with creation 
operators has a counter part with the same $L_0$ eigenvalue obtained by 
acting on $(1 - g) |m_1,m_2,n_1,n_2\rangle$ with the same creation operators;
however, these two states have opposite eigenvalues under $g\in \Z_2$, 
and their contributions cancel in the trace. Thus, only the states 
obtained by acting  
 creation  operators $\alpha^+_{-k}$ or $\bar{\alpha}^+_{-k}$  
on the vacuum $|0,0,0,0\rangle$ will contribute.
Therefore the overall untwisted sector partition function is  
\begin{eqnarray*}
Z_u  &=& 
\frac{1}{2}\left(\frac{1}{|\eta^2|^2}\sum_{ {n_1, m_1 \atop n_2, m_2}
						} 
	q^{\frac{p^2}{2}}\bar{q}^{\frac{\bar{p}^2}{2}}
                      \;  + \;
 \frac { (q\bar q)^{-\frac{1}{12}}}
  {\prod_{n=1}^{\infty}\limits (1 + q^n)^2(1 +\bar{q}^n)^2} \right) Z_{Dirac} 
    \\
 &=& 
  \frac{1}{2}  \left( Z(\tau,\rho) \; + \;
 4\left| \frac{\eta(\sigma)}{ \vartheta_2(\sigma) } \right|^2 
           \right) Z_{Dirac}.
\end{eqnarray*}
Under the symmetry action 
$g$ : $\varphi^+ \rightarrow -\varphi^+ $ the torus has four fixed points. 
This 
implies that there are four twisted ground states with conformal dimension 
$h= \bar h= 1/8$. So one may build four 
distinct Hilbert space sectors. However, these sectors lead
to isomorphic physics, as they are related by translation symmetry of the 
torus. Denote the four twisted sector ground 
states by $|\frac{1}{8},\frac{1}{8}\rangle_f$, where $f=1,2,3,4,$. 
As untwisted bosonic Hilbert space sector, the twisted  
bosonic Hilbert space decomposes into $g = \pm 1$ eigenspaces 
$\widetilde{\mathcal H}_t = 
           \widetilde{\mathcal H}^+_t \oplus \widetilde{\mathcal H}^-_t$ as 
\begin{eqnarray*}
\widetilde{\mathcal H}^+_t &=&  
    \alpha^{+}_{-k_1} \cdot\cdot\cdot \alpha^{+}_{-k_l}
 \bar{\alpha}^+_{-k_{l+1}}\cdot\cdot\cdot\bar{\alpha}^+_{-k_{2j}}
     |\frac{1}{8},\frac{1}{8}\rangle_f \\
\widetilde{\mathcal H}^-_t &=&  
       \alpha^{+}_{-k_1} \cdot\cdot\cdot \alpha^{+}_{-k_l}
 \bar{\alpha}^+_{-k_{l+1}}\cdot\cdot\cdot\bar{\alpha}^+_{-k_{2j + 1}}
     |\frac{1}{8},\frac{1}{8}\rangle_f.
\end{eqnarray*}
where $k_i$ takes half positive integer values. 
By (\ref{gtiw}), the twisted sector partition function is 
\bea\label{2tiw}
Z_t  &=& 
  (q\bar q)^{-\frac{1}{12}} tr_{\widetilde{\mathcal H}_t}{ \frac{1}{2}(1+g) 
                                  q^{L_0}\bar{q}^{\bar L_0}} Z_{Dirac}
\nonumber\\
           &=& 
               4 \times \frac{1}{2}\left(\left| \frac { q^{\frac{1}{24}}}
               {\prod_{n=1}^{\infty}(1 - q^{n-1/2})^2}\right|^2
                      \; +\;
 \left|\frac { q^{\frac{1}{24}}}
 {\prod_{n=1}^{\infty}(1 + q^{n-1/2})^2}\right|^2 \right) Z_{Dirac}
         \nonumber\\
 &=& 
4\times\frac{1}{2}\left(
     \left| \frac{\eta(\sigma)}{ \vartheta_4(\sigma) }\right|^2 \;
               + \; 
\left| \frac{\eta(\sigma)}{ \vartheta_3(\sigma) }\right|^2 
                  \right) Z_{Dirac}.
\eea
Then  the
complete modular invariant $\Z_2$ orbifold partition function has the form
\be\label{z2orb}
Z_{\Z_2-orb} =  \frac{1}{2} \left(Z(\tau, \rho)  
           +
     4\left| \frac{\eta(\sigma)}{\vartheta_2(\sigma)} \right|^2 
                    \; + \; 
     4\left| \frac{\eta(\sigma)}{\vartheta_3(\sigma)} \right|^2
                   \; + \;
    4\left| \frac{\eta(\sigma)}{\vartheta_4(\sigma)} \right|^2 
                   \right)Z_{Dirac}.
\ee
The (c, c), (a, c), and
their complex conjugates, Ramond ground states as well as   
the Witten index
for the $\Z_2$ orbifold 
are the same as those for the $N=2$ theories with $c=3$.
\section{The $\Z_3$ Orbifold }
By dividing the $\Z_3$ symmetry from (\ref{zc3}) for  
$\tau = e^{2\pi i/3}$ and arbitrary $\rho$,
we may construct $\Z_3$ orbifold model. 
The action of $g\in\Z_3$ on the bosonic Hilbert space sectors is given by  
\be\label{z3g}
g|m_1, m_2, n_1, n_2 \rangle =|m_2, -m_1-m_2, n_2-n_1, -n_1 \rangle.
\ee
By (\ref{unorb}), the untwisted sector partition function is 
$$
Z_u  =   
(q\bar q)^{-\frac{1}{8}} tr_{\widetilde{\mathcal H}_u}{\frac{1}{3}(1 + g + g^2)
q^{L_0}\bar{q}^{\bar{L}_0} y^{J_0} \bar{y}^{\bar{J}_0}}.
$$
By taking into account the equations  (\ref{botrans}), (\ref{fertrans})
(\ref{bmod}), (\ref{fmod}) 
and (\ref{z3g}),   
the first term in the trace is equal to the original patition 
function (\ref{zc3}), the second and third term 
receives only contribution from the Hilbert space sector built on 
$|0,0,0,0\rangle$.
The untwisted sector partition function is therefore given by 
\begin{eqnarray*}
Z_u =
\frac{1}{3} \left( Z(\tau = e^{2\pi i/3},\rho,z) \;+\;  
	  \frac{3}{2}\sum_{i = 1}^{4}\left(\left|
			\frac{\vartheta_i (z \;+\; \frac{1}{3}, \sigma)}
			     {\vartheta_1 (\frac{1}{3},\sigma)}\right|^2
\;+\; \left|\frac{\vartheta_i (z - \frac{1}{3}, \sigma)}
             {\vartheta_1 (\frac{1}{3},\sigma)}\right|^2\right)\right).
\end{eqnarray*}
$\Z_3$ does not act freely on the hexagonal torus. Thus one must  consider 
new sectors, the twisted ones.
In the $T^2/{\Z_3}$ $(\tau = e^{2\pi i/3})$ manifold, 
there are three fixed points, and 
one can obtain three Hilbert space sectors corresponding to 
the expansion of the field about each of these fixed points. However these 
three sectors give the same physics. 
The conformal weight of the bosonic twisted 
ground state is $(\frac{1}{9},\frac{1}{9})$. For fermion, twisted sector 
conformal weight is  $(\frac{1}{18},\frac{1}{18})$. Thus the total conformal 
weight of the twisted sector is then  $(\frac{1}{6},\frac{1}{6})$.
States in the twisted sector are generated  by the 
action of creation operators on the twisted ground state. 

By considering  the equations (\ref{botrans}), (\ref{fertrans}),
(\ref{tbmod}), (\ref{tfmod}) and (\ref{z3g}),  
the twisted sector partition function may be written as 
\bea\label{z3tiw}
Z_t &=&  
(q\bar q)^{-\frac{1}{8}} tr_{\widetilde{\mathcal H}_t}
                 {\frac {1}{3}(1 + g + g^2 )
           q^{L_0}\bar{q}^{\bar{L}_0} y^{J_0}\bar{y}^{\bar{J}_0}}\nonumber\\
                             &=&  
3\times\frac{1}{2\times 3} 
		\sum_{i = 1}^{4} 
		\sum_{l=-1}^{1} 
		\left(\left|y^{-\frac{1}{3}}
\frac {\vartheta_i (z + \frac{l}{3} - \frac{\sigma}{3}, \sigma)}
      {\vartheta_1 (\frac{l}{3} - \frac{\sigma}{3}, \sigma)}\right|^2
     			\;+\;
\left|y^{\frac{1}{3}}
	\frac{\vartheta_i (z +\frac{l}{3} +\frac{\sigma}{3},\sigma)}
       {\vartheta_1(\frac{l}{3} + \frac{\sigma}{3},\sigma)}\right|^2\right).
\eea
Then the complete modular invariant 
$\Z_3$ orbifold partition function is 
\bea\label{z3orb}
Z_{\Z_3-orb} & =& \frac{1}{3}
                 \left( Z(\tau = e^{\frac{2\pi i}{3}},\rho,z) 
     	  		\;+\;
 \frac{3}{2}\sum_{i = 1}^{4} \sum_{s = 1}^{2} \left|
		\frac{\vartheta_i (z + \frac{s}{3}, \sigma)}
	             {\vartheta_1 (\frac{s}{3},\sigma)}\right|^2
                        \right.\nonumber\\
& & +\;\left.\frac{3}{2}\sum_{l=-1}^{1}\sum_{i = 1}^{4}\left(
		\left|y^{-\frac{1}{3}}
\frac {\vartheta_i (z + \frac{l}{3} - \frac{\sigma}{3}, \sigma)}
{\vartheta_1 (\frac{l}{3} - \frac{\sigma}{3}, \sigma)} \right|^2
     			\;+\;
\left|y^{\frac{1}{3}}
\frac{\vartheta_i (z +\frac{l}{3} +\frac{\sigma}{3}\sigma)}
     {\vartheta_1(\frac{l}{3} + \frac{\sigma}{3},\sigma)}
                 \right|^2\right)\right).
\eea

We find eight Ramond ground states with conformal dimension 
$(h,\bar h) = (1/8, 1/8)$ and with charges 
$(\pm 1/2, \pm 1/2)$, $3\times(\pm 1/6, \pm 1/6)$, eight NS chiral primary 
states with conformal dimensions  
$(0, 0)$, $(1/2, 1/2)$, $3\times(1/6, 1/6)$, 
$3\times(1/3, 1/3)$  and with charges 
$(0, 0)$, $(1, 1)$, $3\times (1/3, 1/3)$,
$3\times(2/3, 2/3)$, as well as eight NS antichiral primary states having 
the same
conformal dimensions but the opposite charges as the NS chiral fields.
By (\ref{spectralflow}) with $\eta= 1/2$, the ground states of the  Ramond 
sector flow to 
the (c, c) primary states of the NS sector, namely 
\bea\label{nsz3}
\mbox{ \textsl{Ramond ground states} }& \longleftrightarrow &  
             \mbox { \textsl{NS chiral states} }\nonumber\\ 
q^{1/8}\bar{q}^{1/8}y^{-1/2}\bar{y}^{-1/2} & \longleftrightarrow & 
 1\nonumber\\
q^{1/8}\bar{q}^{1/8}y^{1/2}\bar{y}^{1/2} & \longleftrightarrow &
    q^{1/2}\bar{q}^{1/2}y\bar{y}\nonumber\\
3\times q^{1/8}\bar{q}^{1/8}y^{-1/6}\bar{y}^{-1/6} & \longleftrightarrow &
                  3\times q^{1/6}\bar{q}^{1/6}y^{1/3}\bar{y}^{1/3}\nonumber\\
3\times q^{1/8}\bar{q}^{1/8}y^{1/6}\bar{y}^{1/6} & \longleftrightarrow &
             3\times q^{1/3}\bar{q}^{1/3}y^{2/3}\bar{y}^{2/3} .
\eea
(Here $q=e^{2\pi i\sigma}$ and $y = e^{2\pi i z}$.)
If we revers the direction of the spectral flow, 
we get an isomorphism between 
the (a, a) primary states and 
the ground states of the Ramond
sector. By (\ref{R}), (\ref{care}) and (\ref{nsz3}) 
the Witten index and  
the Poincar\'{e} polynomial for the (c,c) states are  
\be\label{wpz3}
Tr(-1)^F = 8,\hspace{4mm}
P(t, \bar t)_{(c,c)} = 
               1 + t \bar t + 3t^{\frac{1}{3}}\bar{t}^{\frac{1}{3}} 
                            + 3t^{\frac{2}{3}}\bar{t}^{\frac{2}{3}}.
\ee
The spectral flow from the NS sector to the NS sector can be obtained by  
flow parameter $\eta = 1$. 
In the spectrum, there are no nontrivial (a, c) or its conjugate
(c, a) states.
\section {The $\Z_4$ Orbifold}
In this section, by dividing the $\Z_4$  symmetry
 from (\ref{zc3}) 
for $\tau =i$ and arbitrary $\rho$, we may formulate 
$\Z_4$ orbifold model.
The action of $g\in\Z_4$
on the bosonic ground state sectors is given by 
\be\label{z4g}
g|m_1, m_2, n_1, n_2\rangle = |m_2,-m_1, n_2, -n_1\rangle.
\ee
Under the rotation group $\Z_4$ the square lattice  has three fixed points. 
An analysis similar to the $\Z_3$ orbifold shows there are twisted sectors 
associated with those fixed points, namely 
one fixed point corresponds to the $\Z_2$ twist and two for the 
$\Z_4$ twist. The weight of the bosonic and fermionic $\Z_4$ twisted ground 
state is $(\frac{3}{32},\frac{3}{32})$ and   
$(\frac{1}{32},\frac{1}{32})$,  
respectively. Thus the  
total conformal weight of the $\Z_4$ twisted sector is then 
$(\frac{1}{8},\frac{1}{8})$.
The total $\Z_4$ orbifold partition function can be obtained  by 
summing over untwisted, $\Z_2$, and $\Z_4$ twisted sectors partition 
functions 
$$
Z_{\Z_4-orb}(\tau =i, \rho, z) = Z_u + Z_{2t} + Z_{4t}.
$$
By (\ref{botrans}), (\ref{fertrans}), 
(\ref{bmod}), (\ref{fmod}), (\ref{z4g}), and (\ref{unorb}),
 we obtain the following untwisted sector partition function 
\begin{eqnarray*}
Z_u & = & 
(q\bar q)^{-\frac{1}{8}} tr_{\widetilde{\mathcal H}_u}
               {\frac {1}{4}(1 + g + g^2 + g^3)
             q^{L_0}\bar{q}^{\bar{L}_0} y^{J_0}\bar{y}^{\bar{J}_0}} 
           \\
& = & 
\frac{1}{4}\left(Z(\tau = i,\rho,z) \;+\; 
                 \sum_{i=1}^{4} \left|\frac{\vartheta_i (z,\sigma)} 
		{\vartheta_2 (\sigma)}\right|^2       
            \;+\;
\sum_{i=1}^{4}\sum_{s=1}^{3}\left|
		\frac{\vartheta_i (z + \frac{s}{4}, \sigma)}
	             {\vartheta_1 (\frac{s}{4},\sigma)}\right|^2 \right).
\end{eqnarray*}
By (\ref{botrans}), (\ref{fertrans}), 
(\ref{tbmod}), (\ref{tfmod}), (\ref{z4g}), and (\ref{gtiw}), 
$\Z_4$ twisted sector partition function may has the form
$$	
 Z_{4t}  =  
\frac{1}{4}\sum_{i,l=1}^{4}\left(
\left|y^{-\frac{1}{4}}
\frac{\vartheta_i(z + \frac{l}{4} - \frac{\sigma}{4}, \sigma)}
     {\vartheta_1(\frac{l}{4} - \frac{\sigma}{4}, \sigma)}\right|^2 
                  \; +\;
\left|y^{\frac{1}{4}}
\frac{\vartheta_i(z + \frac{l}{4} + \frac{\sigma}{4}, \sigma)}
     {\vartheta_1(\frac{l}{4} + \frac{\sigma}{4}, \sigma)}\right|^2 
                  \; +\;
\left|\frac{\vartheta_i(z + \frac{l}{4}, \sigma)}
     {\vartheta_4(\frac{l}{4}, \sigma)}\right|^2\right).
$$
The  $\Z_2$ twisted sector partition function can be read off 
from (\ref{2tiw}) by ommiting the factor of four.
Thus, 
we may  write the modular invariant $\Z_4$ 
orbifold partition function in the following form
\bea\label{z4orb}
Z_{\Z_4-orb} & = & \frac{1}{4}\sum_{i,l=1}^{4}\left(
                Z(\tau =i,\rho,z)
        \;+\;
 \sum_{j=2}^{4}
       \left|\frac{\vartheta_i(z,\sigma)}{\vartheta_j(\sigma)}\right|^2 
                 \; +\; 
\sum_{s=1}^{3}
\left|\frac{\vartheta_i(z + \frac{s}{4}, \sigma)}
            {\vartheta_1(\frac{s}{4}, \sigma)}\right|^2 
                \; +\;
    \right.\nonumber\\
& &         \left.
\left|\frac{\vartheta_i(z + \frac{l}{4}, \sigma)}
            {\vartheta_4(\frac{l}{4}, \sigma)}\right|^2 +
\left|y^{-\frac{1}{4}}\frac
            {\vartheta_i(z + \frac{l}{4} - \frac{\sigma}{4}, \sigma)}
            {\vartheta_1(\frac{l}{4} - \frac{\sigma}{4}, \sigma)}\right|^2 
              \; +\;
       \left|y^{\frac{1}{4}}\frac
            {\vartheta_i(z + \frac{l}{4} + \frac{\sigma}{4}, \sigma)}
        {\vartheta_1(\frac{l}{4} + \frac{\sigma}{4}, \sigma)}\right|^2\right).
\eea

In the spectrum there are  nine Ramond ground states 
which flow to the NS chiral states under the spectral flow operation
(\ref{spectralflow})with flow parameter $\eta = 1/2$
\bea\label{nsz4}
\mbox{ \textsl{Ramond ground states} } & \longleftrightarrow &  
                             \mbox{ \textsl{NS chiral states} }\nonumber\\
	q^{1/8}\bar{q}^{1/8}y^{-1/2}\bar{y}^{-1/2} & \longleftrightarrow 
		& 1\nonumber\\\
	q^{1/8}\bar{q}^{1/8}y^{1/2}\bar{y}^{1/2} & \longleftrightarrow & 
		q^{1/2}\bar{q}^{1/2}y\bar{y}\nonumber\\
2\times q^{1/8}\bar{q}^{1/8}y^{-1/4}\bar{y}^{-1/4} & \longleftrightarrow 
		& 2\times q^{1/8}\bar{q}^{1/8}y^{1/4}\bar{y}^{1/4}\nonumber\\
2\times q^{1/8}\bar{q}^{1/8}y^{1/4}\bar{y}^{1/4} & \longleftrightarrow & 
2 \times q^{3/8}\bar{q}^{3/8} y^{3/4}\bar{y}^{3/4}\nonumber\\
3\times q^{1/8}\bar{q}^{1/8} & \longleftrightarrow & 
3\times q^{1/4} \bar{q}^{1/4}y^{1/2}\bar{y}^{1/2}. 
\eea
There are nine $(a, a)$ states which are given by the 
complex conjugation of $(c, c)$ states.
As in the $\Z_3$ orbifold case, one can get isomorphism between 
the $(a, a)$
primary states and the ground states of Ramond sector by reversing 
the direction of the spectral flow. 
By (\ref{R}), (\ref{care}) and (\ref{nsz4}),  
the Witten index and  
the Poincar\'{e} polynomial for the (c,c) states are  

\bea\label{wpz4}
Tr(-1)^F &=& 9\nonumber\\
P(t, \bar t)_{(c, c)} &=&  
                   1 + t \bar t + 3t^{\frac{1}{2}}\bar{t}^{\frac{1}{2}} 
                              + 
                                  2t^{\frac{1}{4}}\bar{t}^{\frac{1}{4}} 
                              + 
                         2t^{\frac{3}{4}}\bar{t}^{\frac{3}{4}}.\nonumber\\
\eea
With the spectral flow parameter $\eta=1$, the NS sector comes back to 
the NS sector.
One notes that the $\Z_4$ orbifold model contains only (c, c) and their 
conjugate
(a,a) states. For this model, the (a,c) and (c,a) states are trivial and 
consist only of the vacumm state.
\section{The $\Z_6$ Orbifold}
We now construct a $\Z_6$ orbifold model by dividing  $\Z_6$ symmetry 
from (\ref{zc3}) for 
$\tau = e^{2\pi i/3}$ and arbitrary $\rho$.
The bosonic ground state  sectors 
transform as follows under the action of $g\in\Z_6$  
\be
g|m_1, m_2, n_1, n_2\rangle =| m_1 + m_2,-m_1, n_2, -n_1 + n_2\rangle.
\ee
The hexagonal torus has three fixed points under the $\Z_6$ rotation symmetry.
There is a twisted sector associated with each of them. These are 
$\Z_2$, $\Z_3$ and $\Z_6$ twisted sectors. The conformal 
dimension of the 
bosonic and fermionic $\Z_6$ twisted ground state  
is $(\frac{5}{72},\frac{5}{72})$ 
and $(\frac{1}{72},\frac{1}{72})$, respectively.
Thus the total conformal weight of the 
$\Z_6$ twisted ground state is then $(\frac{1}{12},\frac{1}{12})$.
The $\Z_6$ orbifold partition function is the sum of partition functions
of the untwisted, $\Z_2$, $\Z_3$, and $\Z_6$ twisted sectors
\begin{eqnarray*}
Z_{\Z_6-orb}(\tau=e^{\frac{2\pi i}{3}}, \rho,z) = 
                  Z_u + Z_{2t} + Z_{3t} + Z_{6t}.
\end{eqnarray*}
By applying the same method as for the construction of the 
$\Z_2$, $\Z_3$ and $\Z_4$ orbifolds, 
 we obtain the following 
untwisted $\Z_6$ orbifold partition function
\begin{eqnarray*}
Z_u &=&  
(q\bar q)^{-\frac{1}{8}} tr_{\widetilde{\mathcal H}_u}
                          {\frac {1}{6}(1 + g + \cdots + g^5)
           q^{L_0}\bar{q}^{\bar{L}_0} y^{J_0}\bar{y}^{\bar{J}_0}} \\
     &=&
\frac{1}{6}
    \left(Z(\tau=e^{\frac{2\pi i}{3}}, \rho,z)
                     \;\;  + \;\;
\frac{3}{2}\sum_{i=1}^{4} \left| \frac{\vartheta_i (z,\sigma)}
	           {\vartheta_2(\sigma)}\right|^2 \right.\\
          & &  \hspace{4mm}\left.    +\;\;
  \frac{3}{2}\sum_{i=1}^{4}\sum_{s=1}^{2}\left|
          \frac{\vartheta_i (z + \frac{s}{3}, \sigma)}
	       {\vartheta_1 (\frac{s}{3},\sigma)}\right|^2
                   \;\; +	\;\;
\frac{1}{2}\sum_{i=1}^{4}\sum_{l=-1}^{1} \left|
          \frac{\vartheta_i (z + \frac{l}{3}, \sigma)}
	       {\vartheta_2 (\frac{l}{3},\sigma)}\right|^2
 \right).
\end{eqnarray*}
The $\Z_2$ and 
$\Z_3$ twisted sector partition functions can be read off from  
(\ref{2tiw}) and (\ref{z3tiw}) by ommiting the factor of four and three, 
respectively.  
The $\Z_6$ twisted sector partition function may have the form
\begin{eqnarray*}
Z_{6t} &=& 
\frac{1}{12}\sum_{i,k=1}^{4}\sum_{l=-1}^{1}
\left(
      \left|\frac{\vartheta_i(z + \frac{l}{3}, \sigma)}
            {\vartheta_3(\frac{l}{3}, \sigma)}\right|^2 + \;\;
 \left|\frac{\vartheta_i(z + \frac{l}{3}, \sigma)}
            {\vartheta_4(\frac{l}{3}, \sigma)}\right|^2 \right.\\
            & & \left.\;\; + \;\;
\left|y^{\frac{1}{3}}\frac
            {\vartheta_i(z + \frac{l}{3} + \frac{\sigma}{3}, \sigma)}
          {\vartheta_k(\frac{l}{3} + \frac{\sigma}{3}, \sigma)}\right|^2 +\;\;
\left|y^{-\frac{1}{3}}\frac
         {\vartheta_i(z + \frac{l}{3} - \frac{\sigma}{3}, \sigma)}
        {\vartheta_k(\frac{l}{3} - \frac{\sigma}{3}, \sigma)}\right|^2\right).
\end{eqnarray*}
All in all   
 we  obtain the following modular invariant partition function 
\bea\label{z6orb}
Z_{\Z_6-orb} & = & \frac{1}{6}\sum_{i=1}^{4}\sum_{j=2}^{4}
         \left(  Z(\tau=e^{\frac{2\pi i}{3}}, \rho,z) 
                  \;+\;
 \frac{3}{2} \left|
          \frac{\vartheta_i (z,\sigma)}
	       {\vartheta_j(\sigma)} \right|^2
                         \; +\;
  \frac{3}{2} \sum_{s=1}^{2}\left| \frac{\vartheta_i (z + \frac{s}{3}, \sigma)}
	                    {\vartheta_1 (\frac{s}{3},\sigma)} \right|^2 
 \right. \nonumber\\
& & \hspace{10mm} +\;\frac{1}{2}\sum_{l=-1}^{1}\left(
      2\left| y^{-\frac{1}{3}}
         \frac{\vartheta_i (z + \frac{l}{3} - \frac{\sigma}{3}, \sigma)}
              {\vartheta_1 (\frac{l}{3} -\frac{\sigma}{3},\sigma)} \right|^2
\;+\; 
     2\left| y^{\frac{1}{3}}
	\frac{\vartheta_i (z +\frac{l}{3} +\frac{\sigma}{3},\sigma)}
             {\vartheta_1(\frac{l}{3} +\frac{\sigma}{3},\sigma)} \right|^2 
       \right.   \nonumber\\
& & \left.\left. +\;\;
\left| \frac{\vartheta_i (z + \frac{l}{3}, \sigma)}
           {\vartheta_j (\frac{l}{3},\sigma)} \right|^2 
                   \; +\;
\left| y^{-\frac{1}{3}}\frac
            {\vartheta_i(z + \frac{l}{3} - \frac{\sigma}{3}, \sigma)}
            {\vartheta_j(\frac{l}{3} - \frac{\sigma}{3}, \sigma)} \right|^2 
                     \; +\;
\left| y^{\frac{1}{3}}\frac
            {\vartheta_i(z + \frac{l}{3} + \frac{\sigma}{3}, \sigma)}
           {\vartheta_j(\frac{l}{3} + \frac{\sigma}{3}, \sigma)} \right|^2 
      \right)\right).\nonumber\\
\eea

In this model there are 
ten Ramond ground states. Again we  connect the ground 
states of Ramond sector with NS chiral primary states using eq. 
(\ref{spectralflow}) with $\eta= 1/2$ 
\bea\label{nsz6}
	\mbox{ \textsl{Ramond ground states } }& \longleftrightarrow &  
              \mbox{ \textsl{NS chiral states} }
                   \nonumber\\
	q^{1/8}\bar{q}^{1/8}y^{-1/2}\bar{y}^{-1/2} & \longleftrightarrow & 1
                 \nonumber\\
	q^{1/8}\bar{q}^{1/8}y^{1/2}\bar{y}^{1/2} & \longleftrightarrow & 
		q^{1/2}\bar{q}^{1/2}y\bar{y}
                \nonumber\\
2\times q^{1/8}\bar{q}^{1/8} & \longleftrightarrow
& 2\times q^{1/4}\bar{q}^{1/4}y^{1/2}\bar{y}^{1/2}
                       \nonumber\\
        q^{1/8}\bar{q}^{1/8}y^{1/3}\bar{y}^{1/3} & \longleftrightarrow & 
		q^{5/12}\bar{q}^{5/12} y^{5/6}\bar{y}^{5/6}
                      \nonumber\\
	q^{1/8}\bar{q}^{1/8}y^{-1/3}\bar{y}^{-1/3} & \longleftrightarrow &
 q^{1/12}\bar{q}^{1/12}y^{1/6}\bar{y}^{1/6}
                 \nonumber\\
 2\times q^{1/8}\bar{q}^{1/8}y^{1/6}\bar{y}^{1/6} & \longleftrightarrow &
 2\times q^{1/3}\bar{q}^{1/3}y^{2/3}\bar{y}^{2/3}
                 \nonumber\\
 2\times q^{1/8}\bar{q}^{1/8}y^{-1/6}\bar{y}^{-1/6} & \longleftrightarrow &
 2\times q^{1/6}\bar{q}^{1/6}y^{1/3}\bar{y}^{1/3} .
\eea
By (\ref{R}), (\ref{care}) and (\ref{nsz6}), 
The Witten index and the  Poincar\'{e} polynomials
for the (c, c) states are
\bea\label{wpz6}
Tr(-1)^F &=& 10\nonumber\\
P(t, \bar t)_{(c,c)} &=&  
                   1 + t \bar t + 2t^{\frac{1}{2}}\bar{t}^{\frac{1}{2}} 
                              + 
                                  t^{\frac{5}{6}}\bar{t}^{\frac{5}{6}} 
                              + 
                                   t^{\frac{1}{6}}\bar{t}^{\frac{1}{6}} 
                              +
                                  2t^{\frac{2}{3}}\bar{t}^{\frac{2}{3}} 
                               +
                   2t^{\frac{1}{3}}\bar{t}^{\frac{1}{3}}. 
\eea
The (a,a) states are given by the complex conjugation of (c,c) states.
We found no (a, c) and (c,a) states in this model.
\section{$N = 2$ Landau-Ginzburg Model}
In this section, we first review
some of the facts of the 
$N=2$ superconformal Landau-Ginzburg theories by following the articles 
\cite{llw}\cite{vaw}, then we check the spectrum of the (c,c) fields and 
the Witten index.  
The $N=2$ superconformal Landau-Ginzburg  action takes the following form
\be\label{dau}
S = \int d^2zd^4\theta K(\Phi_i, \bar\Phi_i) + \left(\int d^2z d^2\theta 
           W(\Phi_i) + h.c \right).
\ee
%
$\Phi_i$ 
$(i = 1, 2, ...n)$ are the $N=2$ $n$ chiral scalar superfields which satisfy 
the condition
$\bar{D}_{\pm}\Phi_i = D_{\pm}\bar{\Phi}_i = 0$, where
the superderivative defined as 
$D_{\pm} = \frac{\partial }{\partial \theta^\pm} +  
                       \theta^\mp \frac{\partial}{\partial z}$.
The first term (K) is called K\"ahler potential. It includes derivatives of
the superfields. The conformal dimension of those fields is greater than 
$(1,1)$. Such fields are called irrelevant. The second term (W) is called 
superpotential which is a holomorphic function of the superfields. 
It contains only relevant fields, i.e. fields with conformal dimension 
 $(1,1)$ or less than $(1,1)$. The holomorphic 
superpotential $W(\Phi_i)$ is a 
quasi-homogeneous function with isolated singularities at $\Phi_i = 0$. 
In other words  $W(\Phi_i)$  is called 
quasi-homogeneous if it satisfies
\be\label{poten}
W(\lambda^{w_i}\Phi_i) = \lambda^d W(\Phi_i),\hspace{5mm} for \hspace{5mm}
\Phi_i\rightarrow \lambda^{w_i}\Phi_i,
\ee
where $w^i$ and d are integers with no commen factors. 
It has isolated singularity at $\Phi_i = 0$ if it satisfies 
$$ 
W(\Phi_i)|_0 = 0,\hspace{5mm} \partial_i{W(\Phi_j)}|_0 = 0 .
$$

For every isolated quasi-homegeneous superpotential, there exists an $N=2$ 
superconformal field theory. One can read off the $U(1)$ charge of the lowest 
component of the chiral superfields $\Phi_i$ from the action $(\ref{dau})$. 
The $\theta$ integrals in the first term have 
$(left,right)$ charges $(-1,-1)$. 
Because of neutrality of the action  
$W(\Phi_i)$ has charge $(1,1)$. Thus, the chiral superfield 
$\Phi_i$ must 
have charge  $q_i=\frac{w_i}{d}$ for both its left-right moving components.
Now one notes that for any state in the Landau-Ginzburg 
theory $q_L-q_R$ is always an integer. This is true for the chiral superfield 
$\Phi_i$, as it has equal left-right charges. Moreover, it is also 
true for the 
most general fields because they are obtained by taking products of 
$\Phi_i$ and $\bar\Phi_i$, as well as products of their super derivatives.
This implies that one can apply spectral flow to the Landau-Ginzburg models.

The local \textsl{ring} $\mathcal{R}$ of the superpotential 
$W(\Phi_i)$ of the 
Landau-Ginzburg model is 
obtained by taking into account all monomials of chiral superfields $\Phi_i$
and setting $\partial_i{W(\Phi_j)}|_0 = 0$. 
The number of elements of the ring 
is denoted by $\mu =dim\mathcal{R}$. It is called multiplicity of 
 $W(\Phi_i)$. It is also equal to the Witten index $Tr(-1)^F$.

The \textsl{modality} (or moduli is the number of free parameters in the 
theory.) 
m of a quasi-homogeneous superpotentials with isolated 
singularities is given by the number of chiral primary states with charge 
greater than or equal to one.

The \textsl{Poincar\'{e} Polynomial} \cite{llw} for the 
 Landau-Ginzburg theories is
\be\label{poin}
P(t)=Tr_{\mathcal{R}}{t^{d J_0}} =
      \prod_{i=1}^{n}\frac{1-t^{d-w_i}}{1-t^{w_i}},
\hspace{4mm}\mbox{ or } \hspace{4mm}
 P(t,\bar t)=Tr_{\mathcal{R}}{t^{J_0}\bar{t}^{\bar{J}_0}}.     
\ee
This polynomial is only  function of $t\bar t$. 
(Because Landau-Ginzburg primary chiral 
fields have equal left-right charges.) 
For convenience, $t\bar{t}$ is replaced  by the variable $t^d$, where $d$ is 
defined in $(\ref{poten})$. The \textsl{Witten index} \cite{llw} is
\be\label{win}
Tr(-1)^F = P(t=1) = \mu = \prod_{i=1}^{n}\frac{d-w_i}{w_i} = 
                     \prod_{i=1}^{n}\left(\frac{1}{q_i} - 1\right).
\ee
The highest charge and conformal dimension  of 
chiral primary state $|\chi\rangle$ \cite{llw}
are given as 
$$
q_\chi = \sum_{i=1}^{\infty} \frac{d-2w_i}{d} =\sum_i( 1 -2q_i),\hspace{7mm}
h_\chi = \frac{q_{\chi}}{2} 
                  =  
        \sum_{i=1}^{n}\left(\frac{1}{2} - q_i\right).
$$
By using $h_{\chi} = \frac{c}{6}$, 
the central charge of the Landau-Ginzburg   
theory is given as
$$
c= 6h_\chi = 6\sum_{i=1}^{n}\left(\frac{1}{2} - q_i\right).
$$

It is well known \cite{vaw} that 
the \textsl{quasi-homogeneous superpotentials} 
with isolated singularities for 
modality $m=1$ of the Landau-Ginzburg theories at $c=3$ 
are equivalent to the $\Z_M$, $M\in\{3, 4, 6\}$, orbifolds of the $N=2$ 
theories at $c=3$. The corresponding superpotentials are given as 
\bea\label{w3}
W_3(\Phi_1,\Phi_2, \Phi_3) &=& \Phi_1^3 + \Phi_2^3 + \Phi_3^3 
             + 6a\Phi_1\Phi_2\Phi_3, \hspace{2cm} a^3 + 27 \not= 0\\
\label{w4} W_4(\Phi_1,\Phi_2) &=&\Phi_1^4 + \Phi_2^4 + a\Phi_1^2\Phi_2^2,  
         \hspace{4cm} a^2 \not = 4 \\
W_6(\Phi_1,\Phi_2) &=&\Phi_1^3 + \Phi_2^6 + a\Phi_1^2\Phi_2^2,\hspace{35mm}  
               4a^3 + 27 \not= 0.\label{w6}
\eea
%
With the  knowledge in this section, we may write 
the basic linearly independent elements of the (c,c) ring
of superpotential (\ref{w3}) in the following form
\begin{center}
\begin{tabular}{lcccccccc}
chiral fields & 1 & $\Phi_1$ & $\Phi_2$ & $\Phi_3$ & $\Phi_1\Phi_2$  
                 & $\Phi_1\Phi_3$ 
& $\Phi_2\Phi_3$ & $\Phi_1\Phi_2\Phi_3$ \\ 

charges & 0 & 1/3 & 1/3 & 1/3 & 2/3
 & 2/3 & 2/3 & 1\\ 

dimensions & 0 & 1/6 & 1/6 & 1/6 
             & 1/3 & 
1/3 & 1/3 & 1/2.
\\ 
\end{tabular}
\hfill \parbox{1cm}{\be\label{pot} \ee}
\end{center}
By (\ref{win}), (\ref{poin}) and (\ref{pot}), the Witten index 
and Poincar\'{e} polynomial are 
\be\label{wp3}
Tr(-1)^F = 8,
\hspace{4mm}
P(t, \bar t)_{(c,c)} = Tr_{\mathcal R}t^{J_0}\bar{t}^{\bar{J}_0} = 
               1 + t \bar t + 3t^{\frac{1}{3}}\bar{t}^{\frac{1}{3}} 
                            + 3t^{\frac{2}{3}}\bar{t}^{\frac{2}{3}}.
\ee
For the superpotential (\ref{w4}) we have
\begin{center}
\parbox{12cm}{
\begin{tabular}{lccccccccc}
chiral fields & 1 & $\Phi_1$ & $\Phi_2$ & $\Phi_1\Phi_2$  
                 & $\Phi_1^2$ & $\Phi_2^2$ & $\Phi_1^2\Phi_2$ & 
               $\Phi_1\Phi_2^2$ & $\Phi_1^2\Phi_2^2$ \\ 
charges & 0 & 1/4 & 1/4 & 1/2 & 1/2
 & 1/2 & 3/4 & 3/4 & 1 \\ 
dimensions & 0 & 1/8 & 1/8 & 1/4 
             & 1/4 & 1/4 & 3/8 & 3/8 & 1/2  .
\\ 
\end{tabular}
}\hfill \parbox{1cm}{\be\label{tot} \ee}
\end{center}
By (\ref{win}), (\ref{poin}) and (\ref{tot}), the Witten index 
and Poincar\'{e} polynomial are 
\be\label{wp4}
Tr(-1)^F = 9,\hspace{4mm}
P(t, \bar t)_{(c, c)}=  
                   1 + t \bar t + 3t^{\frac{1}{2}}\bar{t}^{\frac{1}{2}} 
                              + 
                                  2t^{\frac{1}{4}}\bar{t}^{\frac{1}{4}} 
                              + 
                         2t^{\frac{3}{4}}\bar{t}^{\frac{3}{4}}
\ee
Similarly, for the  superpotential (\ref{w6}) we may get 
\begin{center}
\parbox{12cm}{
\begin{tabular}{lcccccccccc}

chiral fields & 1 & $\Phi_1$ & $\Phi_2$ & $\Phi_1\Phi_2$ 
                 & $\Phi_2^2$ & $\Phi_2^3$ & $\Phi_2^4$ & $\Phi_1\Phi_2^2$ & 
               $\Phi_1\Phi_2^3$ & $\Phi_1\Phi_2^4$ \\
charges & 0 & 1/6 & 1/3 & 1/2 & 2/3
 & 1/3 & 1/2 & 2/3 & 5/6 
& 1 \\ 
dimensions & 0 & 1/12 & 1/6 & 1/4 
             & 1/3 & 
1/6 & 1/4 & 1/3 & 5/12 & 1/2  .
\\
\end{tabular}
}\hfill \parbox{1cm}{\be\label{kot} \ee}
\end{center}
By (\ref{win}), (\ref{poin}) and (\ref{kot}), the Witten index 
and Poincar\'{e} polynomial are 
\be\label{wp6}
Tr(-1)^F = 10,\hspace{4mm}
P(t, \bar t)_{(c,c)} =  
                   1 + t \bar t + 2t^{\frac{1}{2}}\bar{t}^{\frac{1}{2}} 
                              + 
                                  t^{\frac{5}{6}}\bar{t}^{\frac{5}{6}} 
                              + 
                                   t^{\frac{1}{6}}\bar{t}^{\frac{1}{6}} 
                              +
                                  2t^{\frac{2}{3}}\bar{t}^{\frac{2}{3}} 
                               +
                   2t^{\frac{1}{3}}\bar{t}^{\frac{1}{3}}.
\ee
\subsection*{Conclusion} 
The partition functions for $\Z_M$ orbifolds have been calculated.
The Witten indexes,
the spectrum of (chiral, chiral) fields 
for the $\Z_M$, $M\in\{3,4,6\}$, orbifolds and for the Landau-Ginzburg 
superpotentials (\ref{w3}--\ref{w6})
are given in equations (\ref{wpz3}), (\ref{wpz4}), (\ref{wpz6}), 
(\ref{nsz3}), (\ref{nsz4}), (\ref{nsz6}) and (\ref{pot}), (\ref{tot}), 
(\ref{kot}), (\ref{wp3}), (\ref{wp4}), (\ref{wp6}), respectively.
The results are in in agreement with the 
Landau-Ginzburg predictions of C.~Vafa and 
N.~Warner.
\subsection*{Acknowledgements}
It is great  pleasure to thank my supervisor Professor Werner Nahm 
for countless very helpful and very encouraging discussions. 
I  would like to thank 
K.~Wendland for numerous helpful discussions. I also would like to thank  
 D.~Brungs for his help with Mathematica.
I am grateful to M.~Soika for his help with Latex,
and for proof reading, as well as for his constant hospitality. 
I am also grateful to  H.~Eberle for proof reading.\\
This work was supported by Deutscher Akademischer Austauschdienst (DAAD) 
and in part by TMR.

\end{document}